\documentclass[runningheads]{llncs}
\usepackage[T1]{fontenc}
\usepackage{graphicx}
\usepackage{hyperref}
\usepackage{subcaption}
\usepackage{multirow}
\usepackage{caption}
\usepackage{amssymb}
\usepackage{enumitem}
\usepackage{amsmath}

\begin{document}
\title{Towards Dependable Retrieval-Augmented Generation Using Factual Confidence Prediction}

\author{Florian Geissler\inst{1}\thanks{Email: florian.geissler@iks.fraunhofer.de} \and
Francesco Carella\inst{1} \and
Laura Fieback\inst{2} \and Jakob Spiegelberg \inst{2}}

\institute{Fraunhofer Institute for Cognitive Systems (IKS), Munich, Germany \and
Volkswagen AG, Wolfsburg, Germany}

\authorrunning{F. Geissler et al.}
\titlerunning{Towards Dependable Retrieval-Augmented Generation}

\maketitle

\begin{abstract}
Incorporating specific knowledge into large language models via retrieval-augmented generation (RAG) is a widespread technique that fuels many of today's industry AI applications. A fundamental problem is to assess if the context retrieved by some similarity search provides indeed supporting facts, or instead misguides the generator with irrelevant information.
It is critical to associate meaningful confidence measures about the factuality of the retrieval process with the generated answers.
We present a new, two-staged approach to predict fact faithfulness of the output of retrieval-augmented generations. First, we employ conformal prediction to select only those retrieved chunks who have a high chance to come from the correct source. This approach in itself can improve answer quality by up to $6\%$ in some of the studied datasets, however, the associated statistical guarantees do not hold generally, since the assumption of sample exchangeability depends on the retriever setup. We present diagnostic metrics to assess whether a setup is suitable.
Second, we quantify confidence in the consistency of a generated final answer with a given retrieved context, using an attention-based factuality classifier. This approach can detect inconsistent answers with a chance of up to $77\%$. Our work helps to establish a novel type of certified RAG systems for a broad range of natural language industry applications.
\end{abstract}

\section{Introduction}
\label{sec:intro}

Retrieval augmented generation (RAG) has become a de-facto standard to empower large language models (LLMs) with domain knowledge, such as a curated corpus of source documents. The leap to industry applications has long happened, as frameworks to set up RAG bots in a production environment have become simple and resource-efficient \cite{aws2023,ibmcloud2023}. 
In modern agentic AI applications, RAG is commonly used as a subroutine \cite{Gao2024RAGSurvey}, which creates a dependency of the final outcome on the reliable operation of this mechanism.
Certifiable guarantees of the RAG pipeline must be established, which serve as stepping stones in the safety assurance of the application.

\begin{figure}[h]
\begin{center}
\centerline{\includegraphics[width=1.\textwidth]{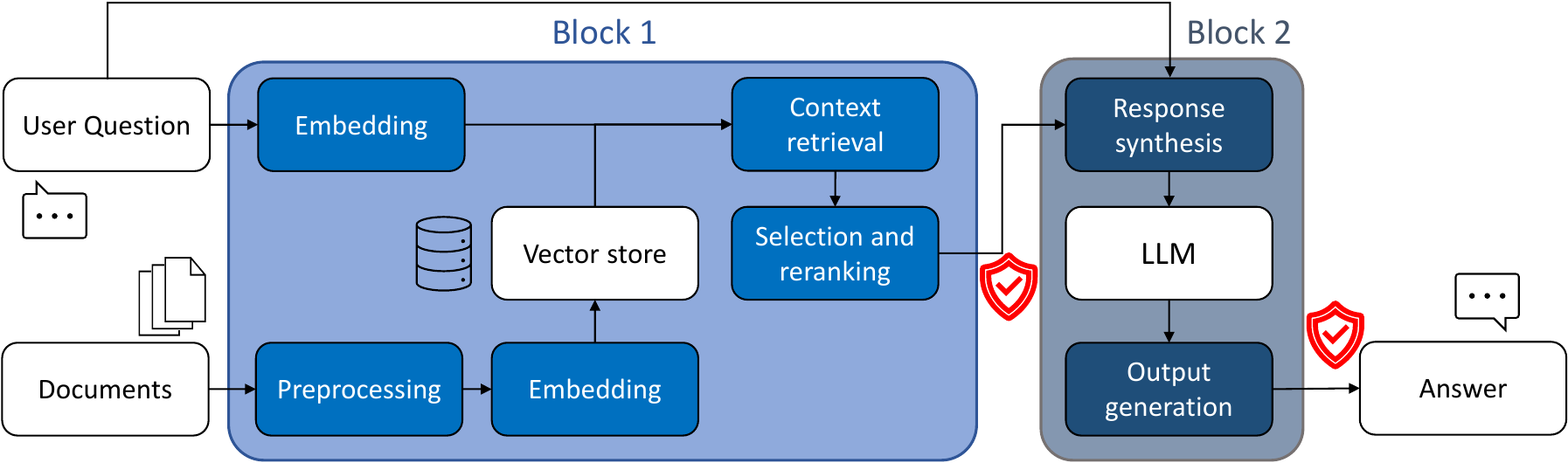}}
\caption{Diagram of a generic RAG workflow, augmented by two certifications as proposed in this article. For the retrieval block 1 (blue), we ensure confidence in retrieving context from the original source. In the generation block 2 (grey), attention monitoring protects the LLM output against potential hallucinations based on certified context.}
\label{fig:system}
\end{center}
\end{figure}

Previous works on reliable RAG systems have focused on extending the basic RAG pipeline - see Fig.~\ref{fig:system} - by various additional mechanisms to improve performance or robustness heuristically, or by analyzing and quantifying failure modes, see Sec.~\ref{sec:related_work}. 
In this work, we pursue instead an architecture-agnostic certification protocol, that does not rely on modifications of individual components but is applicable to any RAG instantiation. 
By conformalizing the retrieval process, we construct certificates for the retrieval quality and analyze the ability of such certificates to generalize (see Sec.~\ref{sec:methods1}). Hallucinated generations from the certified chunks are identified by a detector trained on chunk attentions (Sec.~\ref{sec:methods2}). Our approach is not limited by the choice of retriever or open-source model, and can be combined with any modern, extended RAG workflow. 
In detail, we demonstrate
\begin{itemize}[label={\tiny$\blacksquare$}]
    \item a certifiable RAG pipeline featuring conformalized retrieval and chunk-based hallucination detection, 
    \item an analysis of the pipeline's performance on several data sets. Caveats for real world system designs are discussed,
    \item an analysis of the generalisation capability of the proposed method across various datasets, to validate empirical performance in real-world deployments.
\end{itemize}

\section{Related work}
\label{sec:related_work}
A major barrier for the adoption of LLMs in safety-critical applications is the possibility of factually incorrect answers, so-called hallucinations \cite{azamfirei_large_2023,huang_survey_2025}. 
RAG models ground the available information in external sources \cite{lewis_retrieval-augmented_2020,gao_retrieval-augmented_2024}, but provide no factuality guarantees. Hallucinations originate typically from retrieved context, parametric knowledge, or conflicts between external context and internal knowledge \cite{simhi_constructing_2024}, for which different hallucination detection strategies have been introduced. These methods can be categorized into \textit{closed-book} detection methods, which tackle parametric knowledge hallucinations, and \textit{open-book} methods aiming to detect context hallucinations \cite{simhi_constructing_2024}. While closed-book detection methods rely on internal model representations \cite{azaria_internal_2023,xiao_hallucination_2021} or consistency checks \cite{manakul_selfcheckgpt_2023}, open-book methods measure the importance of context tokens during generation by leveraging attention maps \cite{chuang_lookback_2024} or comparing output distributions with and without external knowledge \cite{lee_refind_2025}.
The factuality of RAG systems can further be improved via advanced processing pipelines, involving additional steps like self-verification \cite{asai2023self}, knowledge consolidation, grading \cite{wang_astute_2025,ge2025conflicting,fang2024hgot}, or uncertainty estimation \cite{perez-utility_2025,li_uncertaintyrag_2024}. Adaptive retrieval and risk-aware generation are strengthened by semantic entropy-guided retrieval \cite{zubkova2025sugar} and counterfactual prompting to promote abstention under low confidence \cite{chen2024controlling}. Recent works also explore how answer reliability can be boosted by agent consensus \cite{chang2025mainrag}. Lastly, security-focused evaluations highlight the need for pipeline-level defenses against adversarial attacks \cite{liang2025saferag}.

\section{Methodology}
\label{sec:methodology}

\subsection{Concept}
\label{sec:concepts}

A generic RAG workflow, as depicted in Fig.~\ref{fig:system}, can be divided into two blocks:
The first block retrieves relevant context from a database, typically a vector store.
We conformalize scores and identify chunks with a statistically guarantee of relevance.
The second block generates an answer for the user question, given the trusted context retrieved in the first block. To confide in the output generated by the LLM, it is key to understand whether the output was obtained based on the provided context and is free of hallucinations. We quantify this using intermediate attention values and a factuality classifier, as explained in Sec.~\ref{sec:methods2}.

\subsection{Retrieval confidence estimation}
\label{sec:methods1}

\begin{figure}[ht]
  \centering
  \begin{subfigure}{0.49\textwidth}
    \centering
    \includegraphics[width=\textwidth]{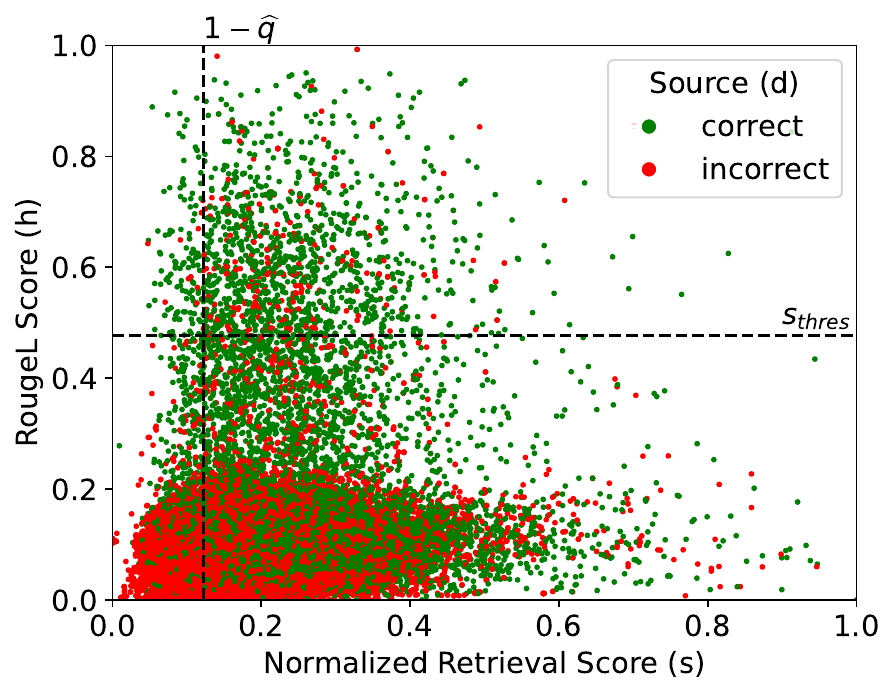}
    \caption{BM25}
    \label{fig:cal1}
  \end{subfigure}\hfill
  \begin{subfigure}{0.49\textwidth}
    \centering
    \includegraphics[width=\textwidth]{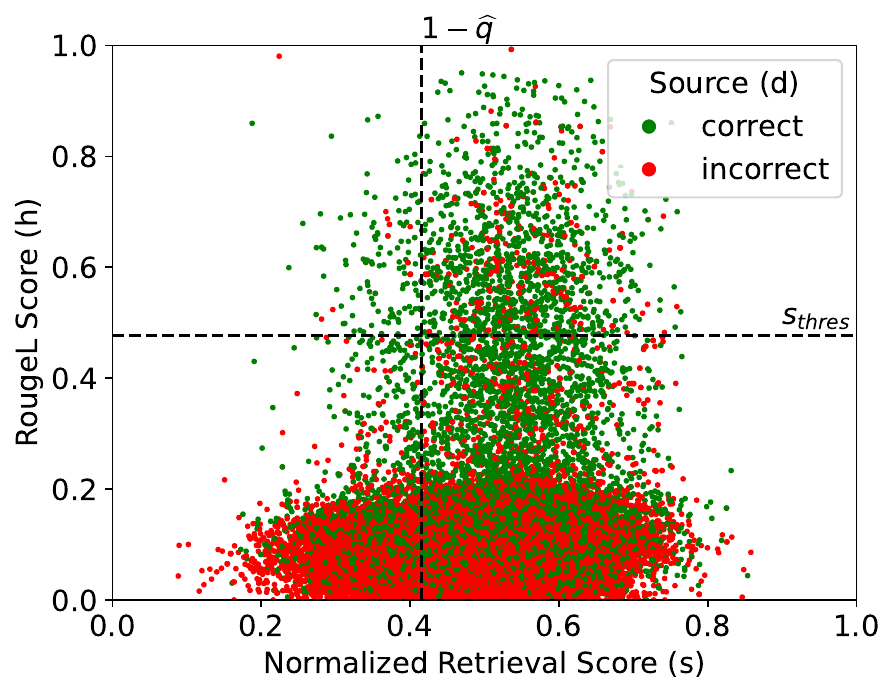}
    \caption{BM25 with cross-encoder reranking}
    \label{fig:cal2}
  \end{subfigure}
  \caption{Characteristics for the BM25 retriever without (a) and with a subsequent reranking using a cross-encoder (b) with $K=10$ and chunk size $512$. Retrieval scores $(s)$ of each chunk are displayed against their rougeL scores $(h)$ and the document source IDs $(d)$, using $5000$ calibration samples of the NQ dataset. 
     For calibration, first, the green chunks with $h > s\textsubscript{thres}$ define a set of correct ground truth, see Eq.~\ref{eq:y_corr}. The retrieval scores of this subset then determine $\widehat{q}$ for a given error rate $\alpha$ (here $0.1$ for illustration), see Eq.~\ref{eq:qhat}. During inference, $\widehat{q}$ serves for conformal prediction of a trust label with statistical guarantees (Eq.~\ref{eq:trusted}).}
  \label{fig:calibration}
\end{figure}

Conformal prediction (CP) is a distribution-free uncertainty quantification method \cite{shafer_tutorial_2008,bates_distribution-free_2021,angelopoulos_gentle_2022} based on the assumption of the exchangeability of data. In a first calibration step, a set of training data is reserved to calculate a non-conformity threshold with a user-given significance level. During inference, this threshold is used to form prediction sets that meet statistical guarantees. We here apply this principle to retrieve document chunks with guaranteed source fidelity.\\
\textbf{Calibration ground truth:} Our datasets consist of triplets of questions $(q)$, reference answers $(r)$ and document indices $d(r)$, where the respective answer is located.
A retriever selects for a given question $i$ the top $K$ candidate chunks, where $K$ is the retrieval depth.
A retrieval score $s$ is assigned to each retrieved chunk $c$, based on the estimated relevance of that chunk for the given query. 
This leads to a set of calibration scores
$S_{cal} = \{ s_{i,j} ;\ i \in (1,\ldots N_{cal}), \ j \in (1, \ldots K)\}$, with $N_{cal}$ being the number of calibration samples.
To facilitate thresholding, we perform a global normalization step to ensure $s\in [0,1]$.
With $n_1 = \min(S_{cal})$ and $n_2 = \max(S_{cal})$, we rescale $S_{cal} \to (S_{cal} - n_1)/(n_2 - n_1)$.
To establish ground truth, we compare a chunk candidate $c$ with the reference answer $r$ for a given question, using a selected similarity metric $h$.
The source document associated with the candidate chunk is $d(c)$.
We first define a score threshold $s\textsubscript{thres}$ as the $\beta$-percentile $P_{\beta}$ of retrieval scores from those chunks that do \textit{not} have the correct document affiliation.  This serves as a minimal scoring threshold for correct chunks,
\begin{equation}
s\textsubscript{thres} = P_{\beta}(\{h(c_{i,j}, r_{i})\ |\ d(c_{i,j}) \neq d(r_{i}) \}).
\label{eq:sthres}
\end{equation}
Depending on the nature of a dataset, the similarity measure ($h$) can be adjusted.
For the NQ dataset, where a source document is rather long, we use a high $\beta=0.99$, while for Ragbench, where source documents are compact, we found that calibration works better by document ID only, i.e. $\beta=0$. 
A candidate chunk is considered \textit{correct} if it passes the threshold \textit{and} has the correct affiliation, leading to boolean chunk ground truth labels
\begin{equation}
   Y_{cal} = \{ h(c_{i,j}, r_{i}) \geq s\textsubscript{thres} \land d(c_{i,j}) = d(r_{i}) \}. \notag
\label{eq:y_corr}
\end{equation}
We take $h$ to be the \textit{RougeL} score, which calculates the longest common subsequence using n-grams of varying length \cite{lin_rouge_2004}.
An example of the distributions of chunk scores and affiliations is given in Fig.~\ref{fig:calibration}. 


\textbf{Prediction threshold:}
During inference, estimates of the correctness of chunk labels are derived with a user-chosen error rate $\alpha \in [0,1]$. Given a calibration set size $n$, we use the adjusted confidence rate, $\delta =\lceil (n+1) (1-\alpha)/n \rceil$, to calculate the $\delta$-percentile, $\widehat{q}$, of correct scores  \cite{angelopoulos_gentle_2022},
\begin{align}
\widehat{q} &=  P_{\delta}(\{(1- s)\ |\ y=true;\ s\in S_{cal}, y\in Y_{cal} \}).
\label{eq:qhat}
\end{align}
%
Connecting to the concept of conformal prediction, we interpret the retriever as a pre-trained classifier function $\widehat{f}$ with a single class (say \textit{trusted}), and the retrieval score of a given chunk, $s = \widehat{f}(c)$ as a normalized prediction of this class. 
If $s\geq 1-\widehat{q}$, the prediction set is the trusted class, otherwise it is empty.
Under the assumption of data sample exchangeability, this gives us the statistical guarantee, that we include the true chunks in the candidate set, with a chance of almost exactly $1-\alpha$ \cite{angelopoulos_gentle_2022},
\begin{equation}
1-\alpha \leq \mathcal{P}(y=true\ |\ s\geq 1-\widehat{q}) \leq 1-\alpha + \frac{1}{n+1},
\label{eq:trusted}
\end{equation}
where $\mathcal{P}$ denotes a probability. We thus obtain a retriever with a $1-\alpha$ guarantee that a chunk scored above the conformality is from the right source, given that sample exchangeability holds.

\textbf{Confidence metrics at inference:}
Let $C$ be the subset of trusted chunks retrieved for a given question,
\begin{equation}
    C = \{c_j\ |\ s_j \geq 1-\widehat{q} ;\ j \in (1, \ldots K)\},
\end{equation}
and $|C|=k \leq K$ the cardinality of this subset.
To estimate the quality of the retriever for a given question, we define the following two metrics
\begin{align}
m_1 &= k > 0, \\
m_2 &= k/K.
\end{align}
In words, $m_1$ is a binary metric stating whether or not there is \textit{at least one} trustworthy chunk among all retrieved chunks, whereas $m_2$ indicates the average portion of trusted chunks among all retrieved chunks.
As we show in Sec.~\ref{sec:results_block1}, the rates $\overline{m_1}$ and $\overline{m_2}$ averaged over a calibration set can serve as diagnostic tools for the suitability of a retriever on a dataset.
\begin{figure*}[h]
\begin{center}
\centerline{\includegraphics[width=1.\textwidth]{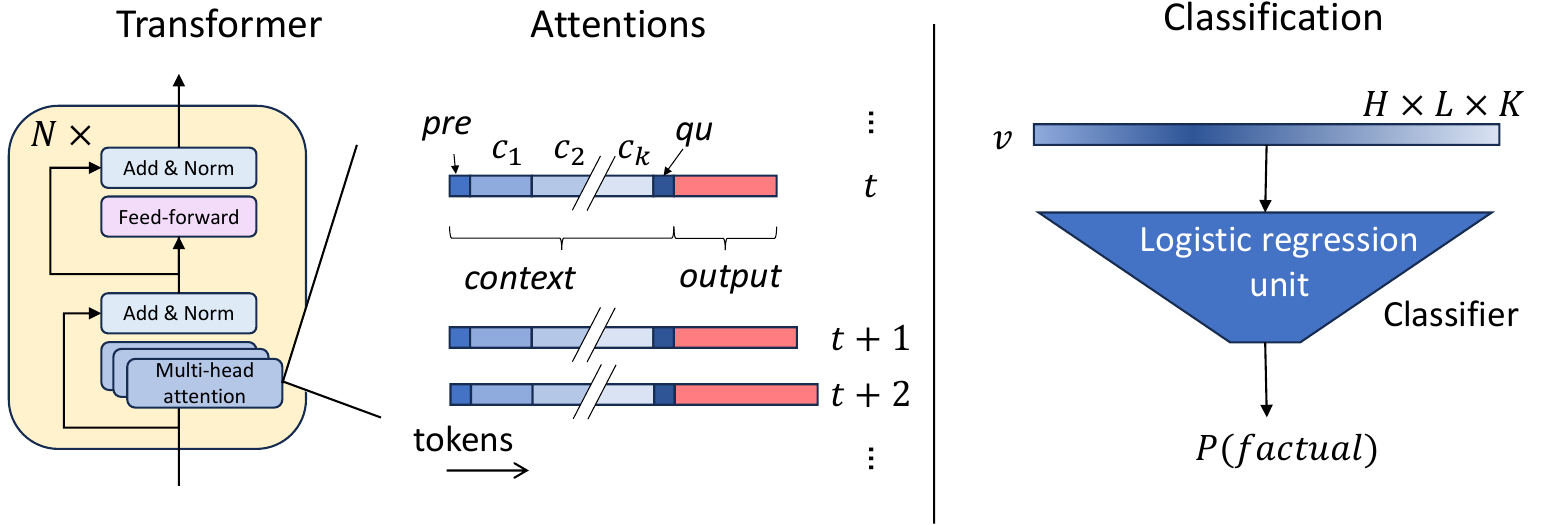}}
\caption{Prediction of factuality, based on the lookback lens concept \cite{chuang_lookback_2024}. Intermediate attention values are extracted with a distillation of the relative attention on individual chunks via lookback ratios. A classifier learns to interpret those ratios to predict whether a generated answer was sufficiently factual.}
\label{fig:block2}
\end{center}
\vskip -0.2in
\end{figure*}

\subsection{Response confidence estimation}
\label{sec:methods2}

\textbf{Attention extraction:}
The attention associated with the LLM output for a given question is a tensor $\alpha^{l, h}_{t, p}$, with indices $l$ for attention layers, $h$ for attention heads, $t \in \{1, \ldots T\}$ runs over the newly generated tokens of the output,
 and $p$ is the considered token index in the prompt. 
 The overall prompt consists of the context and the already generated output tokens (if any), meaning possible segments $x \in \{context, output\}$. The context can again be separated into individual semantic pieces, $context = \{pre, c_1, ..., c_k, qu\}$, where \textit{pre} denotes a system preamble, $c$ a retrieved context chunks with the highest retrieval score first, and \textit{qu} the final user question (see Fig.~\ref{fig:block2}). We track the attentions averaged over tokens in $x$, 
\begin{align}
    A^{l, h}_{t} (x) &= \frac{1}{N_x} \sum_{p \in x} \alpha^{l, h}_{t, p}.
\end{align}
Here, $N_{x}$ is the length of the respective piece $x$ in tokens. The quantity $A$ can be seen as the average attention the model pays to the text block $x$ at a given time step $t$ during generation.

\textbf{Lookback ratios:}
It is then informative to define chunk-wise lookback ratios $LR \in [0,1]$ for a chunk $c_i$ with $i\in \{1,\ldots, K\}$ 
\begin{align}
\label{eq:lr_ratios}
    LR_t^{l,h}(c_i) &= \frac{K\cdot {A}^{l, h}_{t} (c_i)}{{A}^{l, h}_{t} (context) + {A}^{l, h}_{t} (output)}, \\
    LR^{l,h}(c_i) &= \frac{1}{T} \sum_t LR_t^{l,h}(c_i).
    \label{eq:lr_ratios2}
\end{align}  	
Purposefully, the $LR$ here do not include the preamble or the questions in the context, as those parts should not matter for assessment of factuality.
This is to be contrasted with the lookback ratio defined in the original work of \cite{chuang_lookback_2024}, which includes also the static context portions or preamble and question.

\textbf{Factuality classifier:} 
A small classifier is trained to predict hallucinations from those lookback ratios, following \cite{chuang_lookback_2024}. Its inputs are time-averaged lookback ratios (Eq.~\ref{eq:lr_ratios2}) which are unrolled into a single feature vector $v$.
If, for a given input, less than $K$ trusted chunks are identified, we pad $v$ with zeros at those sections, in order to assure a fixed size of $L\times H \times K$. If the full-context lookback ratios are used, this simplifies to $v=[LR^{1,1}(context), \ldots,  LR^{L,H}(context)]$ of size $L\times H$.

The classifier estimates a factuality score $p \in [0,1]$ from the input feature vector $v$, which serves as confidence estimate for the entire answer. 
It is trained using the \textit{answer consistency} ($\gamma_{ac}$) metric defined in \cite{chuang_lookback_2024}, established with a separate LLM judge as ground truth.
A predicted factuality is considered correct if its rounded value matches the provided answer consistency, $\lfloor p \rceil = \gamma_{ac}$. 
Across a dataset, the results are evaluated as the AUROC \cite{pedregosa_scikit-learn_2011} for training and testing. 
In our setup, the classifier architecture consists of a single logistic regression unit \cite{pedregosa_scikit-learn_2011}, see also Fig.~\ref{fig:block2}.

\section{Results}
\label{sec:results}

\subsection{Experimental setup}
\label{sec:setup}

We use the \textit{Natural Questions} (NQ) \cite{kwiatkowski_natural_nodate} dataset by Google as well as the \textit{Ragbench} (RB) \cite{friel_ragbench_2025} collection which in turn contains $12$ datasets of question-and-answer pairs from different domains.
For each dataset, a random selection of the \textit{train} split is chosen, with a maximum cap of $10K$ samples for practical feasibility. 
Half of those samples are used for calibration, half for testing.
As a representative of a popular family of open-source model architectures, we select Llama-3-8B-Instruct \cite{grattafiori_llama_2024,wolf_huggingfaces_2020}, 
while GPT3.5 \cite{brown_language_2020} is used as LLM judge. 
To learn factuality classification, we split the lookback ratio data in a ratio of $3:2$ for training and validation of the classifier, respectively.\\ 
Furthermore, the \textit{text-ada-embedding-002-2} embedding model \cite{openai_openai_nodate} is used with a chunk size of $512$. For retrieval, we deploy the BM25 retriever from the \textit{Llama-index} library \cite{noauthor_llamaindex_nodate}, while experiments with reranking as described in Sec.~\ref{sec:methods1} make use of the \textit{stsb-roberta-base} pair-wise cross-encoder from the sentence transformer library \cite{noauthor_sentence_nodate}. Note that we want to examine certification in RAG, not strive for state of the art performance. Throughout the experiments in Sec.~\ref{sec:results}, we use $K=10$ and $\alpha=0.1$.

For the classification of lookback ratios, we train a simple logistic regression unit from \textit{sklearn} \cite{pedregosa_scikit-learn_2011}.
As hyperparameters, we set a maximum iteration of $1000$ and class weight \textit{balanced}. 
The lookback ratios are normalized by linear scaling to a range of $[0,1]$ for each individual data sample.

\begin{figure}[ht]
  \centering
  \begin{subfigure}{0.49\textwidth}
    \centering
    \includegraphics[width=\textwidth]{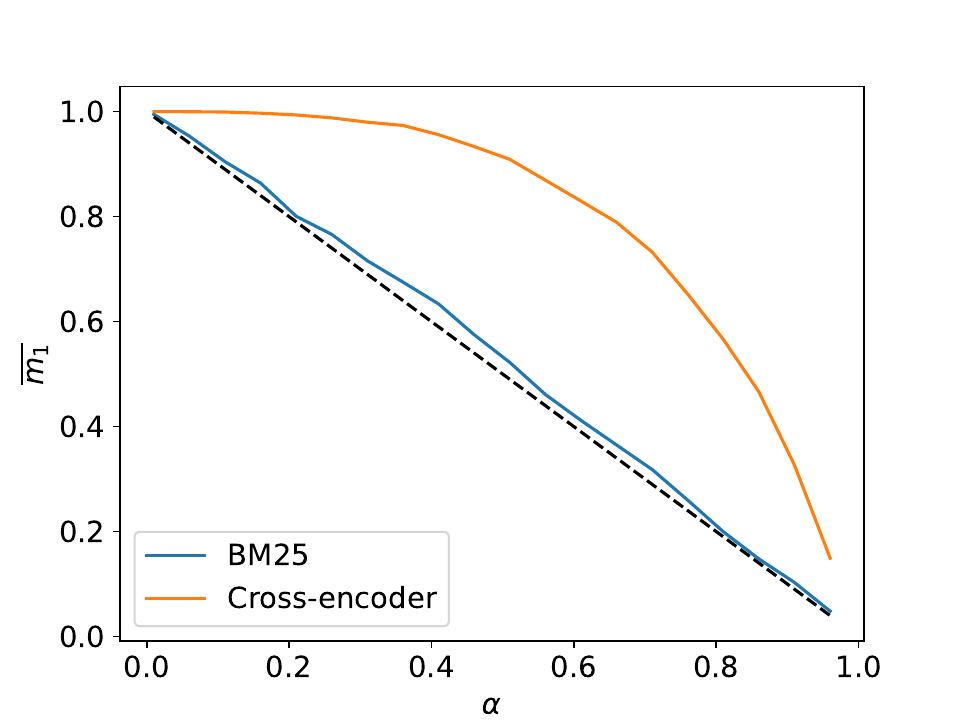}
  \end{subfigure}\hfill
  \begin{subfigure}{0.49\textwidth}
    \centering
    \includegraphics[width=\textwidth]{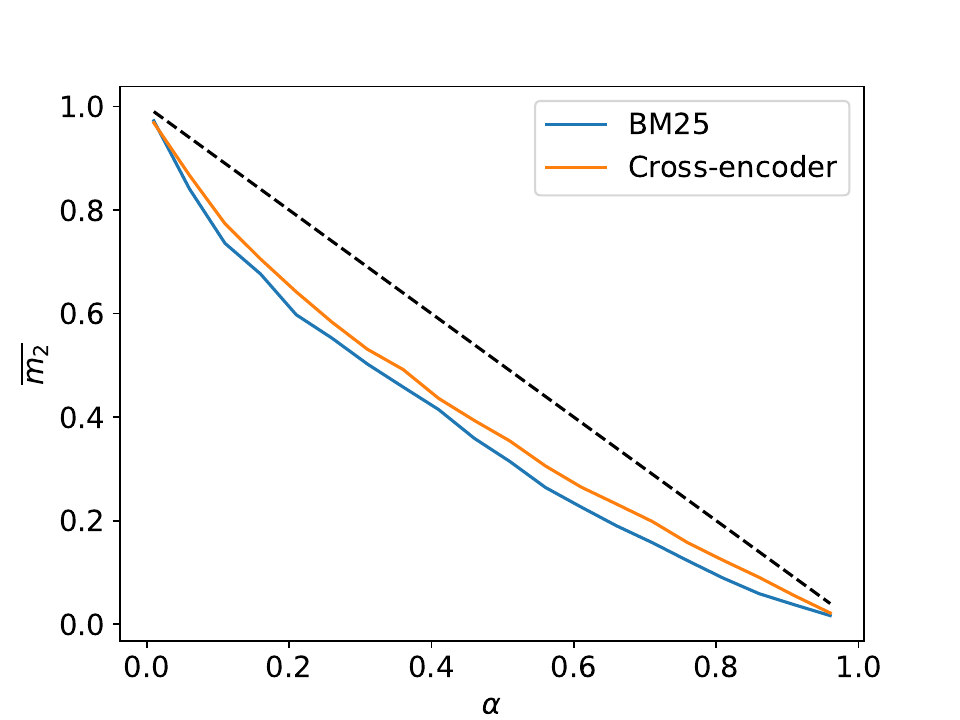}
  \end{subfigure}
\caption{The $m_1$ and $m_2$ metrics averaged over a calibration set under variation of the error rate $\alpha$ for the BM25 retriever, with and without subsequent reranking with the cross-encoder. We used $5000$ samples of NQ and $K=10$.}
    \label{fig:m1m2}
\end{figure}

\subsection{Retrieval confidence estimation}
\label{sec:results_block1}

\textbf{Sample exchangeability:} The suitability of a retriever for a given setup can be evaluated by the average $m_1$ and $m_2$ metrics defined in Sec.~\ref{sec:methods1}.
The retrieval score of an arbitrary, \textit{correct} (according to ground truth) chunk is - by construction of $\widehat{q}$ in Eq.~\ref{eq:qhat}, assuming full exchangeability of samples - above the trust threshold with a chance of $1-\alpha$, and given that there exists at least one correct source chunk for each question.
Importantly, however, the retrieval depth $K$ is always finite in practice. The assumption of exchangeability of samples is only given within the pool of the respective top $K$ chunks. 
If the retriever is struggling to find the correct document source, and instead assigns high retrieval scores only to incorrect chunks, correct chunks may not appear anymore in the top-$K$, meaning that they are under-represented and $m_1$ will fall short of the expectation $1-\alpha$. On the other hand, if the retriever assigns high retrieval scores only to correct source chunks and finds so many of them that incorrect chunks are cut off from the top $K$, almost all retrieved content will be accepted and $m_1$ is unexpectedly high.
As a consequence, the bounds in Eq.~\ref{eq:trusted} do not hold strictly, but approximately, if the top-$K$ sample pool is sufficiently representative of correct chunks.

Tab.~\ref{tab:results_m1m2} shows the average $m_1$ and $m_2$ values across the studied datasets.
We see that most results are close to the expected bounds of conformal prediction (see Eq.~\ref{eq:trusted}).
Notable exceptions are \textit{RB-tatqa} and \textit{RB-finqa}. Both datasets contain financial questions and the BM25 retriever fails to identify the correct chunk in the top $K$: Almost always either zero or a maximum amount of $K$ chunks ($\mathbf{\overline{m_2}}\approx \mathbf{\overline{m_1}}$) is retrieved, indicating its struggle to assign a meaningful score. Since calibration scores are evaluated over true chunks in the calibration data, absence of true chunks in the top $K$ breaks exchangeability of calibration data and test data. This is a crucial pitfall to consider when designing practical application using CP. 

To further elaborate on this important observation, we study the variation of $m_1$, $m_2$ with the significance level $\alpha$ for different ranking techniques. An example using NQ data is shown in Fig.~\ref{fig:m1m2}.
We see that the BM25 retriever characteristic is close to the expectation of exchangeability. 
With reranking, we see in Fig.~\ref{fig:m1m2} that the acceptance of correct chunk increases significantly. The balance of correct and incorrect chunks gets skewed towards more correct chunks with higher scores. Unfortunately, however, this result is overly optimistic at finite retrieval depth, as incorrect source chunks get cut off alltogether. 
We choose the base retriever without cross-encoder reranking for the further experiments of Sec.~\ref{sec:results}.
%

\begin{table}[ht]
    \centering
    \begin{subtable}{0.48\textwidth}
    \centering
        \begin{tabular}{|c|p{1.5cm}|p{1.5cm}|}
            \hline
            \centering
            \textbf{Dataset} & $\mathbf{\overline{m_1}} (\%)$ &  $\mathbf{\overline{m_2}} (\%)$ \\
            \hline \hline
            NQ & $ 91.1 \pm 0.4$ & $ 74.9\pm 0.5$ \\
            \hline
            RB-covidqa & $ 90.4 \pm 1.2 $ & $ 66.4 \pm 1.5 $ \\
            \hline
            RB-cuad & $ 94.8 \pm 0.8$ & $ 94.8 \pm 0.8$ \\
            \hline
            RB-delucionqa & $ 89.3 \pm 1.6$ & $ 77.5\pm 1.9$ \\
            \hline
            RB-emanual & $ 82.7\pm 1.6$ & $ 77.8\pm 1.7$ \\
            \hline
            RB-expertqa & $ 93.7 \pm 0.9 $ & $ 66.9\pm 1.3$ \\
            \hline
            RB-finqa & $ 75.8\pm 0.6$ & $ 72.6\pm 0.6$ \\
            \hline
            RB-hagrid & $ 92.4\pm 0.7$ & $ 40.7\pm 0.8$ \\
            \hline
            RB-hotpotqa & $ 97.9\pm 0.5$ & $ 27.0\pm 0.9$ \\
            \hline
            RB-msmarco & $ 94.9\pm 0.7$ & $ 37.9\pm 1.0$ \\
            \hline
            RB-pubmedqa & $ 91.3\pm 0.4$ & $ 59.2\pm 0.5$ \\
            \hline
            RB-tatqa & $ 61.4 \pm 0.7$ & $ 61.4\pm 0.7$ \\
            \hline
            RB-techqa & $ 93.8\pm 1.0$ & $ 87.5\pm 1.2 $ \\
            \hline
        \end{tabular}
    \caption{Averaged $m_1$ and $m_2$ characteristic of the retriever on the studied datasets, with standard error. Parameters: $\alpha=0.1$, $K=10$. For RB we used $\beta=0$, for NQ $\beta=0.99$, as explained in the text.}
    \label{tab:results_m1m2}
  \end{subtable}\hfill
  \begin{subtable}{0.48\textwidth}
    \centering
    \begin{tabular}{|c|p{1.5cm}|p{1.5cm}|}
            \hline
            \centering
            \multirow{2}{*}{\textbf{Dataset}} & \multicolumn{2}{|c|}{$\mathbf{\overline{\gamma_{ac}}}$} \\ \cline{2-3}
            & filter & no filter \\
            \hline \hline
            NQ & $ 65.1\pm 0.7$ & $ 65.3\pm 0.7$\\
            \hline
            RB-covidqa & $ 73.9 \pm 1.8$ & $72.3 \pm 1.8$\\
            \hline
            RB-cuad & $ 57.1 \pm 1.9 $ & $51.6 \pm 1.9$ \\
            \hline
            RB-delucionqa & $77.7 \pm 2.2$ &  $77.6 \pm 2.2$\\
            \hline
            RB-emanual & $ 69.2 \pm 2.0$ & $78.4 \pm 1.8$\\
            \hline
            RB-expertqa & $ 59.5 \pm 1.7$ & $53.0 \pm 1.8$\\
            \hline
            RB-finqa & $ 49.5\pm 0.7$ & $ 59.7\pm 0.7$\\
            \hline
            RB-hagrid & $ 74.1 \pm 1.2$ & $ 70.0 \pm 1.2$\\
            \hline
            RB-hotpotqa & $ 78.8 \pm 1.3$ & $74.5 \pm 1.4$\\
            \hline
            RB-msmarco & $ 74.1\pm 1.4$ & $68.8 \pm 1.5$\\
            \hline
            RB-pubmedqa & $ 78.6\pm 0.6$ &  $77.9\pm 0.6$\\
            \hline
            RB-tatqa & $ 49.7\pm 0.7$ & $ 65.3\pm 0.7 $\\
            \hline
            RB-techqa & $ 50.9\pm 1.2 $ & $ 49.3 \pm 2.1$\\
            \hline
        \end{tabular}
        \caption{Averaged answer consistency $\mathbf{\overline{\gamma_{ac}}}$ for Llama3 with and without chunk filtering based on confidence estimation. With filtering, untrusted chunks are removed to for the subsequent LLM call.}
        \label{tab:results_filter_or_not}
  \end{subtable}
  \caption{Caption}
  \label{tab:results}
\end{table}

\begin{figure}[h]
\begin{center}
\centerline{\includegraphics[width=0.65\textwidth]{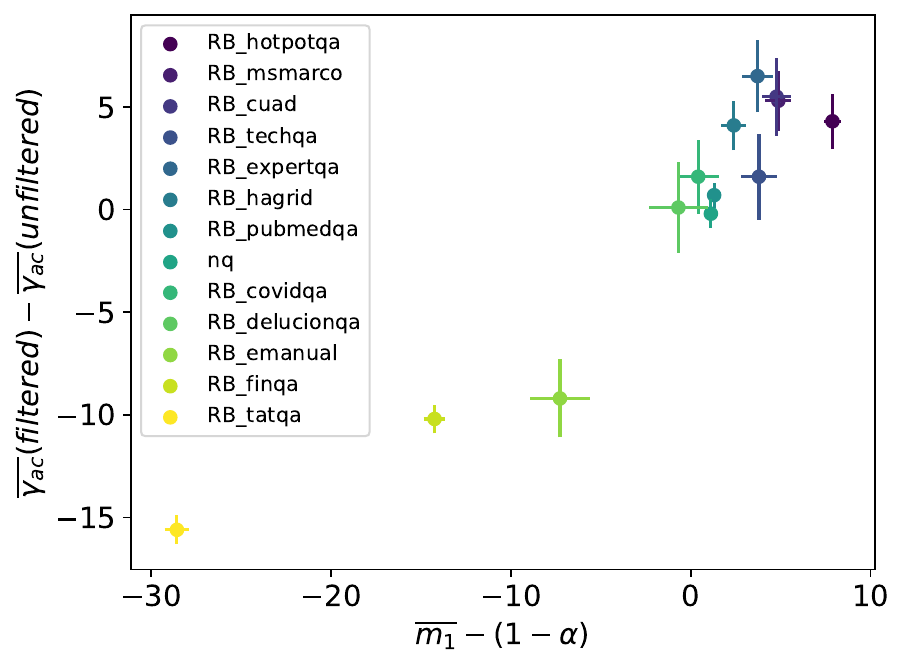}}
\caption{Retriever performance on a dataset, as measured by deviation of the average $m_1$ from the expected $1-\alpha$, versus the impact of chunk filtering on the consistency of the subsequent LLM output, as given in Tabs.~\ref{tab:results_m1m2}-\ref{tab:results_filter_or_not}. We see that chunk filtering is beneficial if the retriever performance is high, and vice versa.}
\label{fig:m_vs_gac}
\end{center}
\end{figure}

\textbf{Impact of chunk filtering:} We next evaluate, whether the filtering for trusted chunks directly affects the quality of LLM answers during RAG. Results for LLama-3 are shown in Tab.~\ref{tab:results_filter_or_not} and Fig.~\ref{fig:m_vs_gac}, where the answer consistency is measured as $\gamma_{ac}$.
We observe variations between the data sets. 
For \textit{RB-expertqa}, for example, a boost in answer consistency of more than $6\%$ emerges. On the other hand, \textit{RB-finqa} shows the opposite effect, and context filtering actually degrades the answer consistency by roughly $10\%$. 
More generally, the change in answer consistency performance is correlated with the $m_1$ retriever performance, as illustrated in Fig.~\ref{fig:m_vs_gac}.
We here visualize the difference from the CP expectation, $\mathbf{\overline{m_1}}-(1-\alpha)$, against the delta in answer quality. A low retriever performance in this metric leads to a negative effect of chunk filtering and vice versa. This is due to the fact that correct chunk sources are no longer strictly associated with high retrieval scores. In our experiments, this lacking retriever performance is found for \textit{RB-finqa}, \textit{RB-tatqa}, and \textit{RB-emanual}. For strong retrievers, chunk filtering benefits the answer quality of the LLMs positively, i.e., $\mathbf{\overline{m_1}}\geq 1-\alpha$.

\begin{figure}[ht]
  \centering
  \begin{subfigure}{0.49\textwidth}
    \centering
    \includegraphics[width=\textwidth]{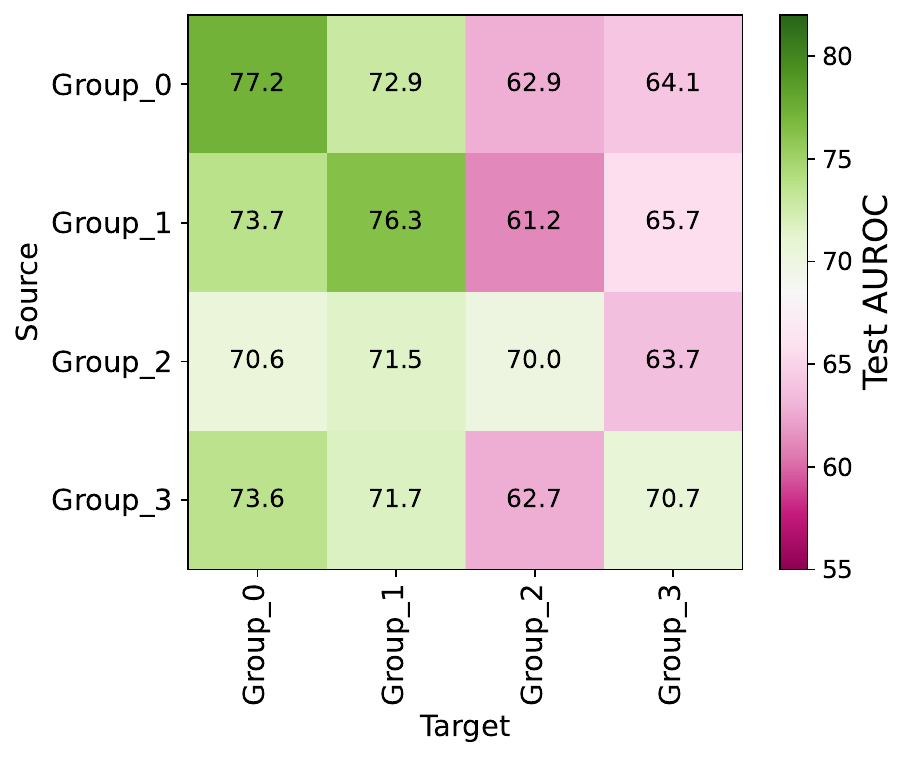}
    \caption{CW LR}
    \label{fig:chunks}
  \end{subfigure}\hfill
  \begin{subfigure}{0.49\textwidth}
    \centering
    \includegraphics[width=\textwidth]{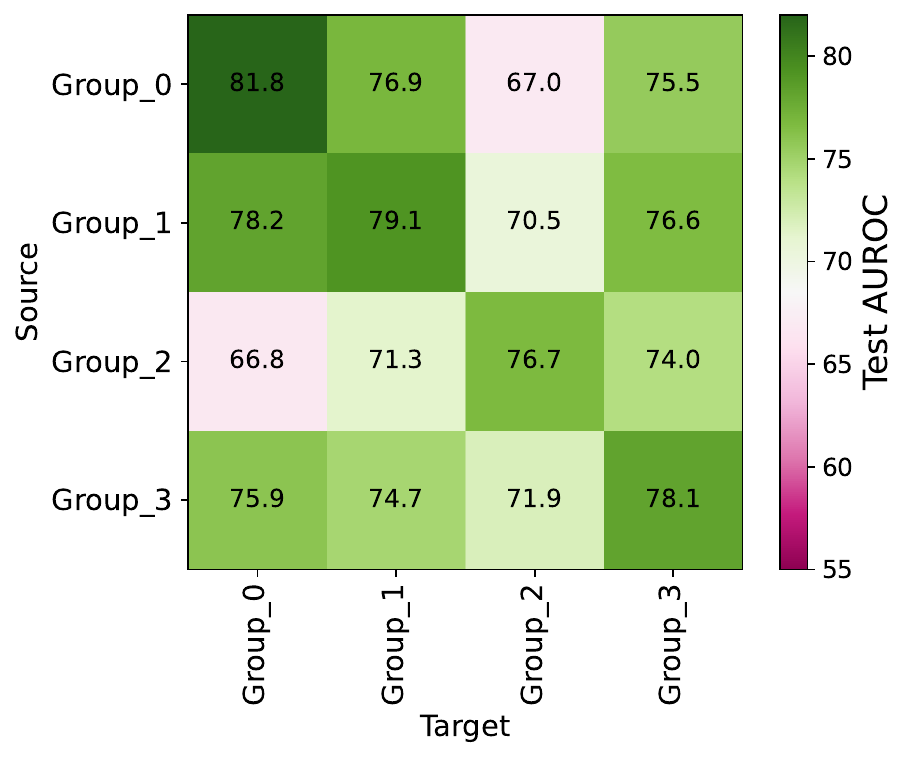}
    \caption{FC LR}
    \label{fig:context}
  \end{subfigure}
\caption{Test AUROCs for classifiers trained on the training split of the \textit{source} dataset, and tested on the test portion of the \textit{target} dataset. Off-diagonal elements represent cross-validation scenarios. We compare classifiers trained on chunk-wise (CW) $LR$ (\textit{left}) and lookback ratios using the average full context (FC) (\textit{right}) as in \cite{chuang_lookback_2024}.}
\label{fig:test_heatmap}
\end{figure}


We conclude that poor retrieval not only breaks CP guarantees, it also degrades subsequent generation performance.
The $m_1$ and $m_2$ metrics devised in this article therefore serve as a metric to characterize the suitability of a RAG setup for a given dataset and retrieval depth. If such a suitability is established, test time confidence guarantees from CP can be constructed.

\subsection{Response confidence estimation}
\label{sec:results_block2}

In this section, we derive a confidence estimation for the factuality of a generated LLM answer, given a previously retrieved context (Sec.~\ref{sec:methods2} or block 2 of the RAG pipeline). To that end, we train a classifier of answer consistency.
Since the individual datasets vary significantly in magnitude, we chose to group them in clusters of comparable size (see Tab.~\ref{tab:pacc_training}) with group 0: (RB-tatqa, RB-expertqa, RB-cuad), group 1: (RB-finqa, RB-msmarco, RB-techqa, RB-delucionqa) , group 2:  (RB-pubmedqa, RB-hotpotqa, RB-covidqa), and group 3: (NQ, RB-hagrid, RB-emanual).
This also ensures that hallucination detection does not learn patterns that are too specific to one given dataset.\\
As classifier input, we compare chunk-wise (CW) lookback ratios and those averaged across the full context (FC). The latter importantly includes system preamble and question, see Fig.~\ref{fig:block2} and discussion in Sec.~\ref{sec:methods2}.
From Tab.~\ref{tab:pacc_training}, we first observe that both strategies result in classifiers with similar training performance for each group. Fig.~\ref{fig:test_heatmap} shows that AUROCs are highest for validation within the same dataset group - up to $77\%$ and $82\%$, respectively - while cross-validation with other groups exhibit performance drops of up to $16\%$.

Interestingly, we observe that the FC LR classifiers outperform the CW LR ones by up to $12\%$, especially in cross-validation.
As an intermediate experiment, we average out the chunk dimension of the lookback ratios in Eq.~\ref{eq:lr_ratios}, capturing all context attention except from preamble and question in the feature vector. The resulting classifiers show some individual improvements in off-diagonal validation, yet do not lead to any test AUROC $>77\%$, hence still fall short of the classifiers trained on FC LR, where the only difference is the inclusion of prompt and question attention.

Our study in this section hence reveals the following key finding: 
The superior performance of the full-context hallucination detection over the one based on chunk context only (Fig.~\ref{fig:test_heatmap}) is attributed to relevant attention on the system prompt and question. This implies that hallucination patterns are partially learned from generic information that is unrelated to factual correctness. For example, the classifier might learn that questions from the domain of finance are more likely to elicit incorrect answers. 
Full-context hallucination detectors can thus be misleading, instead we propose to use CW LR in reliable RAG systems despite the lower test AUROC for confidence estimation, as non-retrieval related sources of attention are excluded by design. Therefore, our predictor provides a better factual grounding.
Our result may inspire further systematic studies on the impact of domain-specific system prompts and questions on model attentions in the future. 

\begin{table}[t]
    \centering
    \begin{tabular}{|c|c|c|c|}
            \hline
            \multirow{2}{*}{Group ID} & \multirow{2}{*}{Size} & \multicolumn{2}{|c|}{Train AUROC} \\ \cline{3-4}
             & & CW LR & FC LR \\
            \hline \hline
            0 & \centering 6447 & $83.7$& $85.8$\\
            \hline
            1 & \centering 6810 & $86.5$& $85.0$\\
            \hline
            2 & \centering 6463 & $87.5$ &$84.1$\\
            \hline
            3 & \centering 6840 & $86.3$&$83.6$\\
            \hline
        \end{tabular}
        \vspace{0.2cm}
    \caption{Training AUROC of dataset groups, using clusters of similar size for Chunk-wise (CW) and Full-context (FC) LR.}
    \label{tab:pacc_training}
\end{table}


\section{Conclusion}
\label{sec:conclusion}
We present a comprehensive workflow to quantify factual confidence in RAG systems.
Our approach provides separate metrics for the stages of context retrieval and LLM-based answer generation, respectively, and therefore allows to pinpoint potential weaknesses in practical RAG pipelines. We demonstrate that conformal prediction is an effective way to provide confidence guarantees for the correctness of chunks, but also showcase that the assumption of exchangeability does not apply a priori to retrieval scores at finite retrieval depth, such that combinations of retrievers and datasets have to be chosen with care. The cardinalities of the trusted chunk sets serve as an indicator metric for the suitability of such combinations. To predict factuality of LLM answers from a given context, we train a classifier on the chunk-wise lookback ratio. This custom quantity represents the attention the model spends on the respective context chunks during answer generation and leads to high hallucination detection rates. In contrast to previously proposed work on lookback ratios, our approach excludes attentions on prompt segments unrelated to retrieval, which can possibly have misleading effects on the classifier. 
Our findings guide the development towards dependable RAG applications in safety-critical systems.

\section*{Acknowledgment}
Parts of this work have been funded by the Free State of Bavaria in the DSgenAI project (Grant Nr.: RMF-SG20-3410-2-18-4). The results, opinions and conclusions expressed in this publication are not necessarily those of Volkswagen Aktiengesellschaft.

\bibliographystyle{splncs04.bst}
\bibliography{bibliography_reliable_rag_shortened}

@article{Gao2024RAGSurvey,
  author = {Gao, Yunfan and Xiong, Yun and Gao, Xinyu and Jia, Kangxiang and Pan, Jinliu and Bi, Yuxi and Dai, Yi and Sun, Jiawei and Wang, Meng and Wang, Haofen},
  title = {Retrieval-Augmented Generation for Large Language Models: A Survey},
  journal = {arXiv preprint arXiv:2312.10997},
  year = {2024},
  url = {https://arxiv.org/abs/2312.10997v5},
  eprint = {2312.10997},
  archiveprefix = {arXiv},
  primaryclass = {cs.CL},
}

@misc{angelopoulos_gentle_2022,
 author = {Angelopoulos, Anastasios N. and Bates, Stephen},
 doi = {10.48550/arXiv.2107.07511},
 language = {en},
 month = {December},
 note = {arXiv:2107.07511 [cs]},
 publisher = {arXiv},
 title = {A {Gentle} {Introduction} to {Conformal} {Prediction} and {Distribution}-{Free} {Uncertainty} {Quantification}},
 url = {http://arxiv.org/abs/2107.07511},
 urldate = {2025-03-14},
 year = {2022}
}

@inproceedings{asai2023self,
 author = {Asai, Akari and Wu, Zeqiu and Wang, Yizhong and others},
 booktitle = {The Twelfth International Conference on Learning Representations},
 title = {Self-rag: Learning to retrieve, generate, and critique through self-reflection},
 year = {2023}
}

@misc{aws2023,
 author = {{Amazon Web Services}},
 howpublished = {\url{https://docs.aws.amazon.com/prescriptive-guidance/latest/patterns/develop-advanced-generative-ai-chat-based}-\url{assistants-by-using-rag-and-react-prompting.html}},
 note = {Accessed: 2023-10-01},
 title = {Develop Advanced Generative AI Chat-Based Assistants by Using RAG and React Prompting},
 year = {2023}
}

@article{azamfirei_large_2023,
 author = {Azamfirei, Razvan and Kudchadkar, Sapna R. and Fackler, James},
 doi = {10.1186/s13054-023-04393-x},
 issn = {1364-8535},
 journal = {Critical Care},
 language = {en},
 month = {March},
 number = {1},
 pages = {120},
 title = {Large language models and the perils of their hallucinations},
 url = {https://ccforum.biomedcentral.com/articles/10.1186/s13054-023-04393-x},
 urldate = {2025-04-22},
 volume = {27},
 year = {2023}
}

@inproceedings{azaria_internal_2023,
 address = {Singapore},
 author = {Azaria, Amos and Mitchell, Tom},
 booktitle = {Findings of the {Association} for {Computational} {Linguistics}: {EMNLP} 2023},
 doi = {10.18653/v1/2023.findings-emnlp.68},
 language = {en},
 pages = {967--976},
 publisher = {Association for Computational Linguistics},
 title = {The {Internal} {State} of an {LLM} {Knows} {When} {It}’s {Lying}},
 url = {https://aclanthology.org/2023.findings-emnlp.68},
 urldate = {2025-04-22},
 year = {2023}
}

@article{bates_distribution-free_2021,
 author = {Bates, Stephen and Angelopoulos, Anastasios and Lei, Lihua and others},
 doi = {10.1145/3478535},
 issn = {0004-5411, 1557-735X},
 journal = {Journal of the ACM},
 language = {en},
 month = {December},
 number = {6},
 pages = {1--34},
 title = {Distribution-free, {Risk}-controlling {Prediction} {Sets}},
 url = {https://dl.acm.org/doi/10.1145/3478535},
 urldate = {2025-07-03},
 volume = {68},
 year = {2021}
}

@misc{brown_language_2020,
 author = {Brown, Tom B. and Mann, Benjamin and Ryder, Nick and others},
 doi = {10.48550/arXiv.2005.14165},
 month = {July},
 note = {arXiv:2005.14165 [cs]},
 publisher = {arXiv},
 title = {Language {Models} are {Few}-{Shot} {Learners}},
 url = {http://arxiv.org/abs/2005.14165},
 urldate = {2025-04-30},
 year = {2020}
}

@inproceedings{chang2025mainrag,
 author = {Chang, Chia-Yuan and Jiang, Zhimeng and Rakesh, Vineeth and others},
 booktitle = {Proceedings of the 63rd Annual Meeting of the ACL},
 title = {{MAIN-RAG}: Multi-Agent Filtering Retrieval-Augmented Generation},
 year = {2025}
}

@inproceedings{chen2024controlling,
 author = {Chen, Lu and Zhang, Ruqing and Guo, Jiafeng and others},
 booktitle = {Findings of the Association for Computational Linguistics: EMNLP 2024},
 doi = {10.18653/v1/2024.findings-emnlp.133},
 month = {Nov},
 note = {Code \& RC-RAG benchmark available},
 publisher = {Association for Computational Linguistics},
 title = {Controlling Risk of Retrieval-Augmented Generation: A Counterfactual Prompting Framework},
 year = {2024}
}

@misc{chuang_lookback_2024,
 author = {Chuang, Yung-Sung and Qiu, Linlu and Hsieh, Cheng-Yu and others},
 doi = {10.48550/arXiv.2407.07071},
 language = {en},
 month = {October},
 note = {arXiv:2407.07071 [cs]},
 publisher = {arXiv},
 shorttitle = {Lookback {Lens}},
 title = {Lookback {Lens}: {Detecting} and {Mitigating} {Contextual} {Hallucinations} in {Large} {Language} {Models} {Using} {Only} {Attention} {Maps}},
 url = {http://arxiv.org/abs/2407.07071},
 urldate = {2025-03-14},
 year = {2024}
}

@inproceedings{fang2024hgot,
 author = {Fang, Yihao and Thomas, Stephen and Zhu, Xiaodan},
 booktitle = {Proceedings of the 4th Workshop on Trustworthy NLP (TrustNLP@ACL)},
 pages = {118--144},
 title = {{HGOT}: Hierarchical Graph of Thoughts for Retrieval-Augmented In-Context Learning in Factuality Evaluation},
 year = {2024}
}

@misc{friel_ragbench_2025,
 author = {Friel, Robert and Belyi, Masha and Sanyal, Atindriyo},
 doi = {10.48550/arXiv.2407.11005},
 month = {January},
 note = {http://arxiv.org/abs/2407.11005},
 publisher = {arXiv},
 shorttitle = {{RAGBench}},
 title = {{RAGBench}: {Explainable} {Benchmark} for {Retrieval}-{Augmented} {Generation} {Systems}},
 url = {http://arxiv.org/abs/2407.11005},
 urldate = {2025-04-30},
 year = {2025}
}

@misc{gao_retrieval-augmented_2024,
 author = {Gao, Yunfan and Xiong, Yun and Gao, Xinyu and others},
 doi = {10.48550/arXiv.2312.10997},
 month = {March},
 note = {arXiv:2312.10997},
 publisher = {arXiv},
 shorttitle = {Retrieval-{Augmented} {Generation} for {Large} {Language} {Models}},
 title = {Retrieval-{Augmented} {Generation} for {Large} {Language} {Models}: {A} {Survey}},
 url = {http://arxiv.org/abs/2312.10997},
 urldate = {2025-04-22},
 year = {2024}
}

@article{ge2025conflicting,
 author = {Ge, Ziyu and Wu, Yuhao and Chin, Daniel Wai Kit and others},
 journal = {Proc. of IJCAI-25},
 title = {Resolving Conflicting Evidence in Automated Fact-Checking: A Study on Retrieval-Augmented LLMs},
 year = {2025}
}

@misc{grattafiori_llama_2024,
 author = {Grattafiori, Aaron and Dubey, Abhimanyu and Jauhri, Abhinav and others},
 doi = {10.48550/arXiv.2407.21783},
 month = {November},
 note = {arXiv:2407.21783},
 publisher = {arXiv},
 title = {The {Llama} 3 {Herd} of {Models}},
 url = {http://arxiv.org/abs/2407.21783},
 urldate = {2025-04-22},
 year = {2024}
}

@article{huang_survey_2025,
 author = {Huang, Lei and Yu, Weijiang and Ma, Weitao and others},
 doi = {10.1145/3703155},
 issn = {1046-8188, 1558-2868},
 journal = {ACM Transactions on Information Systems},
 language = {en},
 month = {March},
 number = {2},
 pages = {1--55},
 shorttitle = {A {Survey} on {Hallucination} in {Large} {Language} {Models}},
 title = {A {Survey} on {Hallucination} in {Large} {Language} {Models}: {Principles}, {Taxonomy}, {Challenges}, and {Open} {Questions}},
 url = {https://dl.acm.org/doi/10.1145/3703155},
 urldate = {2025-04-22},
 volume = {43},
 year = {2025}
}

@misc{ibmcloud2023,
 author = {{IBM Developer}},
 howpublished = {\url{https://developer.ibm.com/articles/awb-scenarios-options-for-rag-da/}},
 note = {Accessed: 2023-10-01},
 title = {AWB Scenarios: Options for RAG DA},
 year = {2023}
}

@article{kwiatkowski_natural_nodate,
 author = {Kwiatkowski, Tom and Palomaki, Jennimaria and Redfield, Olivia and others},
 language = {en},
 title = {Natural {Questions}: {A} {Benchmark} for {Question} {Answering} {Research}}
}

@misc{lee_refind_2025,
 author = {Lee, DongGeon and Yu, Hwanjo},
 doi = {10.48550/arXiv.2502.13622},
 month = {April},
 note = {arXiv:2502.13622},
 publisher = {arXiv},
 shorttitle = {{REFIND} at {SemEval}-2025 {Task} 3},
 title = {{REFIND} at {SemEval}-2025 {Task} 3: {Retrieval}-{Augmented} {Factuality} {Hallucination} {Detection} in {Large} {Language} {Models}},
 url = {http://arxiv.org/abs/2502.13622},
 urldate = {2025-04-22},
 year = {2025}
}

@inproceedings{lewis_retrieval-augmented_2020,
 author = {Lewis, Patrick and Perez, Ethan and Piktus, Aleksandra and others},
 booktitle = {Advances in {Neural} {Information} {Processing} {Systems}},
 pages = {9459--9474},
 publisher = {Curran Associates, Inc.},
 title = {Retrieval-{Augmented} {Generation} for {Knowledge}-{Intensive} {NLP} {Tasks}},
 url = {https://proceedings.neurips.cc/paper/2020/hash/6b493230205f780e1bc26945df7481e5-Abstract.html},
 urldate = {2025-04-22},
 volume = {33},
 year = {2020}
}

@inproceedings{li_uncertaintyrag_2024,
 author = {Li, Zixuan and Xiong, Jing and Ye, Fanghua and others},
 booktitle = {arXiv preprint arXiv:2410.02719},
 month = {Oct},
 title = {UncertaintyRAG: Span‐Level Uncertainty Enhanced Long‐Context Modeling for Retrieval‐Augmented Generation},
 year = {2024}
}

@article{liang2025saferag,
 author = {Liang, Xun and Niu, Simin and Li, Zhiyu and others},
 journal = {arXiv:2501.18636},
 title = {{SafeRAG}: Benchmarking Security in Retrieval-Augmented Generation of Large Language Models},
 year = {2025}
}

@inproceedings{lin_rouge_2004,
 address = {Barcelona, Spain},
 author = {Lin, Chin-Yew},
 booktitle = {Text {Summarization} {Branches} {Out}},
 month = {July},
 pages = {74--81},
 publisher = {Association for Computational Linguistics},
 title = {{ROUGE}: {A} {Package} for {Automatic} {Evaluation} of {Summaries}},
 url = {https://aclanthology.org/W04-1013/},
 year = {2004}
}

@inproceedings{manakul_selfcheckgpt_2023,
 address = {Singapore},
 author = {Manakul, Potsawee and Liusie, Adian and Gales, Mark},
 booktitle = {Proceedings of the 2023 {Conference} on {Empirical} {Methods} in {Natural} {Language} {Processing}},
 doi = {10.18653/v1/2023.emnlp-main.557},
 language = {en},
 pages = {9004--9017},
 publisher = {Association for Computational Linguistics},
 shorttitle = {{SelfCheckGPT}},
 title = {{SelfCheckGPT}: {Zero}-{Resource} {Black}-{Box} {Hallucination} {Detection} for {Generative} {Large} {Language} {Models}},
 url = {https://aclanthology.org/2023.emnlp-main.557},
 urldate = {2025-04-22},
 year = {2023}
}

@misc{noauthor_llamaindex_nodate,
 author = {LLamaIndex Contributors},
 note = {Accessed: 2023-10-03},
 title = {LLamaIndex: A Data Framework for LLMs},
 url = {https://github.com/jerryjliu/llama_index},
 year = {2023}
}

@misc{noauthor_sentence_nodate,
 author = {Nils Reimers and Iryna Gurevych},
 note = {Accessed: 2023-10-03},
 title = {Sentence Transformers: Multilingual Sentence and Image Embeddings},
 url = {https://www.sbert.net},
 year = {2019}
}

@misc{openai_openai_nodate,
 author = {{OpenAI}},
 title = {{OpenAI} {Embeddings}},
 url = {https://platform.openai.com/docs/guides/embeddings}
}

@article{pedregosa_scikit-learn_2011,
 author = {Pedregosa, Fabian and Varoquaux, Gaël and Gramfort, Alexandre and others},
 journal = {Journal of Machine Learning Research},
 number = {85},
 pages = {2825--2830},
 title = {Scikit-learn: {Machine} {Learning} in {Python}},
 url = {http://jmlr.org/papers/v12/pedregosa11a.html},
 volume = {12},
 year = {2011}
}

@inproceedings{perez-utility_2025,
 author = {Perez‐Beltrachini, Laura and Lapata, Mirella},
 journal = {arXiv preprint arXiv:2502.18108},
 month = {Feb},
 note = {Code available at \url{https://github.com/lauhaide/ragu}},
 title = {Uncertainty Quantification in Retrieval Augmented Question Answering},
 year = {2025}
}

@article{shafer_tutorial_2008,
 author = {Shafer, Glenn and Vovk, Vladimir},
 issn = {1533-7928},
 journal = {Journal of Machine Learning Research},
 number = {12},
 pages = {371--421},
 title = {A {Tutorial} on {Conformal} {Prediction}},
 url = {http://jmlr.org/papers/v9/shafer08a.html},
 urldate = {2025-07-03},
 volume = {9},
 year = {2008}
}

@misc{simhi_constructing_2024,
 author = {Simhi, Adi and Herzig, Jonathan and Szpektor, Idan and others},
 doi = {10.48550/arXiv.2404.09971},
 month = {July},
 note = {arXiv:2404.09971},
 publisher = {arXiv},
 title = {Constructing {Benchmarks} and {Interventions} for {Combating} {Hallucinations} in {LLMs}},
 url = {http://arxiv.org/abs/2404.09971},
 urldate = {2025-04-22},
 year = {2024}
}

@inproceedings{wang_astute_2025,
 author = {Wang, Fei and Wan, Xingchen and Sun, Ruoxi and others},
 journal = {arXiv preprint arXiv:2410.07176},
 title = {{ASTUTE RAG: Overcoming Imperfect Retrieval Augmentation and Knowledge Conflicts for Large Language Models}},
 url = {https://arxiv.org/abs/2410.07176},
 year = {2025}
}

@misc{wolf_huggingfaces_2020,
 author = {Wolf, Thomas and Debut, Lysandre and Sanh, Victor and others},
 doi = {10.48550/arXiv.1910.03771},
 month = {July},
 note = {arXiv:1910.03771 [cs]},
 publisher = {arXiv},
 shorttitle = {{HuggingFace}'s {Transformers}},
 title = {{HuggingFace}'s {Transformers}: {State}-of-the-art {Natural} {Language} {Processing}},
 url = {http://arxiv.org/abs/1910.03771},
 urldate = {2025-04-08},
 year = {2020}
}

@inproceedings{xiao_hallucination_2021,
 address = {Online},
 author = {Xiao, Yijun and Wang, William Yang},
 booktitle = {Proceedings of the 16th {Conference} of the {European} {Chapter} of the {Association} for {Computational} {Linguistics}: {Main} {Volume}},
 doi = {10.18653/v1/2021.eacl-main.236},
 language = {en},
 pages = {2734--2744},
 publisher = {Association for Computational Linguistics},
 title = {On {Hallucination} and {Predictive} {Uncertainty} in {Conditional} {Language} {Generation}},
 url = {https://aclanthology.org/2021.eacl-main.236},
 urldate = {2025-04-22},
 year = {2021}
}

@inproceedings{zubkova2025sugar,
 author = {Zubkova, Hanna and Park, Ji-Hoon and Lee, Seong-Whan},
 journal = {arXiv preprint arXiv:2501.04899},
 month = {Jan},
 note = {Semantic entropy guides adaptive retrieval},
 title = {SUGAR: Leveraging Contextual Confidence for Smarter Retrieval},
 year = {2025}
}

\end{document}